\documentclass[11pt,onecolumn,draftcls]{IEEEtran}

\usepackage{latexsym}
\usepackage{mdwmath}
\usepackage{amsmath}
\usepackage{amssymb}

\usepackage{color}
\usepackage{psfrag}
\usepackage{alltt}
\usepackage{moreverb}

\newtheorem{rem}{Remark}
\newtheorem{theo}{Theorem}

\newtheorem{lem}{Lemma}

\newcommand{\tr}{\text{tr}}
\newcommand{\rank}{\text{rank}}

\newcommand{\eqdef}{\stackrel{\triangle}{=}}

\usepackage{graphicx}
\usepackage{graphics}
\usepackage{color}
\usepackage{psfrag}
\usepackage{pstricks, pst-node, pst-tree}
\usepackage{pstricks-add}

\begin{document}
\author{Dimitris S. Papailiopoulos and Alexandros G. Dimakis\footnote{The authors are with the Department of Electrical Engineering, University of Southern California, Los Angeles, CA 90089. Email:\texttt{\{papailio, dimakis\}@usc.edu}}\footnote{A preliminary version of this work was presented in \cite{PD}.}}

\title{Interference Alignment as a Rank Constrained Rank Minimization}
\maketitle

\begin{abstract}
We show that the maximization of the sum degrees-of-freedom for the static flat-fading multiple-input multiple-output (MIMO) interference channel is equivalent to a rank constrained rank minimization  problem (RCRM), when the signal spaces span all available dimensions.
The rank minimization corresponds to maximizing interference alignment (IA) so that interference spans the lowest dimensional subspace possible.
The rank constraints account for the useful signal spaces spanning all available spatial dimensions. 
That way, we reformulate all IA requirements to requirements involving ranks.
Then, we present a convex relaxation of the RCRM problem inspired by recent results in compressed sensing and low-rank matrix completion theory that rely on approximating rank with the nuclear norm. 
We show that the convex envelope of the sum of ranks of the interference matrices is the normalized sum of their corresponding nuclear norms and introduce tractable constraints that are asymptotically equivalent to the rank constraints for the initial problem.
We also show that our heuristic relaxation can be tuned for the multi-cell interference channel.
Furthermore, we experimentally show that in many cases the proposed algorithm attains perfect interference alignment and in some cases outperforms previous approaches for finding precoding and zero-forcing matrices for interference alignment. 
\end{abstract}

\section{Introduction}

A recent information-theoretic breakthrough established that at the high signal-to-noise (SNR) regime every user in a $K$-user wireless interference network can enjoy half the capacity of the interference free case \cite{Cadambe1}.
Therefore, interference is not a fundamental limitation for such networks since it accounts for only constant scaling of the interference free case capacity, provided that it is sufficiently mitigated.
Such a surprising result is possible when interference alignment is employed. IA is a sophisticated technique first presented in \cite{Ali1} and subsequently utilized in \cite{Cadambe1} as a means of showing the achievability of $\frac{K}{2}$ degrees-of-freedom (DoF) for the $K$-user interference channel. 
The DoF of an interference channel can be interpreted as the number of interference free signaling dimensions, including time, frequency, or space.

Intuitively, IA serves as a means for obtaining as many interference free dimensions for communication as possible and in practice stands for designing the transmit and receive strategies for each user-receiver pair of a wireless network \cite{Cadambe1}-\cite{Gou}. 
For the case of static flat-fading MIMO channels, such as the ones studied in \cite{Jafar1} and \cite{Jafar2}, where both transmitter and receiver have perfect channel knowledge, the flexibility is confined to designing the transmit precoding and receive zero-forcing matrices to maximize the achievable {\it spatial} DoF. 
Unfortunately, such matrices are NP-hard to obtain \cite{Luo}, closed form solutions have been found only for a few special cases such as \cite{Cadambe1} and \cite{Roland}, and the problem is open for limited dimensions \cite{Yetis}. 
Even characterizing the feasibility of perfect IA is a highly non-trivial task as discussed in recent work~\cite{Yetis} and \cite{Luo}. 
The hardness to either find perfect IA solutions or even decide for feasibility is the cost of the problem's over constrained nature. 
A review of the current status of IA techniques is presented in \cite{Cadambe3}. 

As an alternative to closed form designs, several algorithmic approaches have been proposed in the literature such as \cite{Gomadam}, \cite{Peters1}, \cite{Peters2}, and \cite{Sung1}. 
Many of those methods aim to minimize the {\it interference leakage} at each receiver so that -at best case- interference alignment is perfectly attained. 
The suggested insight for their effectiveness is that when {\it perfect} interference alignment is possible, then interference leakage will be zero and such algorithms may obtain the optimal solutions. 
Although a fair metric to optimize, as we show in this paper, interference leakage is not the tightest approximation to the notion of DoF which is typically the desired objective.
When perfect alignment of interferences is not attained, the objective remains to maximize the available spatial DoF, that is the prelog factor of the capacity at the high-SNR regime. 

In this work, we present a variational characterazation of IA in terms of signal and interference space ranks. 
Specifically, we pose full rank constraints on the useful signal spaces and minimize the rank of the interference spaces.
The full rank constraints ensure useful signal spaces spanning all available spatial dimensions. The rank minimization guarantees interference spaces collapsing to the smallest dimensional subspaces possible. 
We show that under the full rank constraints, minimizing the sum of ranks of the interference matrices is equivalent to maximizing the sum of spatial DoF for static flat-fading MIMO systems.
The variational characterization, even though it is harder to solve exactly than solving a set of bilinear equations for perfect IA as the ones in \cite{Gomadam} and \cite{Yetis}, suggests a natural relaxation that leads to a good approximation algorithm.

Hence, we establish a new heuristic for near optimal interference alignment and maximization of the sum of spatial DoF. 
Using results from \cite{Recht} and \cite{Candes} 
we 
suggest that the sum of the nuclear norms of the interference matrices is the best objective function in terms of convex functions and is equivalent to the $\ell_1$-norm of the singular values of the interference matrices. 
As intuition suggests, the $\ell_1$-norm minimization of the singular values of interference matrices will provide sparse solutions, translating to more interference free signaling dimensions. 
Interestingly, we show that the leakage minimization techniques presented in \cite{Gomadam}, \cite{Peters1}, and \cite{Peters2} minimize the $\ell_2$-norm of the singular values of the interference matrices which accounts for ``low'' energy solutions rather than sparse ones. 
To deal with the (non-convex) full rank constraints we suggest that positivity constraints on minimum eigenvalues of positive definite matrices serve as a well motivated approximation.

An interesting aspect of our approach is its inherent robustness to a challenging caveat that the leakage minimization approaches face: for some cases of special interest, such as symbol extended single antenna interference channels, the iterative leakage minimization approaches may converge to zero dimensional, or rank deficient signal spaces, yielding zero or very low DoF \cite{Gomadam2}.
This caveat is due to the fact that the beamforming and zeroforcing matrices for IA are constructed as a function only of the interference links. Then,  the randomness of the wireless medium is left to take care of  the full rank signal space conditions, which it does when the channel matrices are full and do not have a specific structure.
Our nuclear norm heuristic avoids these singularities due to positivity constraints on eigenvalues, explicitly enforced on all signal spaces; these constraints involve all direct links of the network.
Therefore, our approach can be used to generate interference alignment beamforming and zeroforcing matrices for any channel structure of interest.

Then, we extend our approximation algorithm to the $K$-cell interference channel \cite{Suh}, where each cell consists of several users and show that additional affine constraints on the precoding matrices are posed.
These affine constraints are tractable and can be added in a straightforward manner to our approximation, hence, we develop a similar convex programming approach.
This approximation is once more motivated by the fact that IA can be posed as a rank constrained rank minimization on an interference channel with structural constraints on the beamforming matrices.

The last section contains an experimental evaluation of the proposed algorithm. 
Our experiments suggest that the proposed scheme is sum-DoF optimal for many setups where perfect IA is possible. 
Furthermore, in some cases it provides extra DoF compared to the leakage minimization approach, when perfect IA is not attained and is robust to singularities met by the leakage minimization approach on diagonal channel structures.
We conclude with a discussion on further research directions. 

\section{System Model}
We consider static flat-fading $K$-user MIMO interference wireless systems consisting of  $K$ transmitters and $K$ receivers. 
We assume that each transmitting user is equipped with $M_{\text{t}}$ transmit antennas, each receiver with $M_{\text{r}}$ receive antennas, and all $K$ transmitting users are synchronizing their transmissions. 
Each user, say $k\in\mathcal{K}$, wishes to communicate a symbol vector ${\bf x}_k\in\mathbb{C}^{d\times 1}$ to its associated receiver, where $\mathcal{K}\eqdef\{1,2,\ldots,K\}$ and $d$ represents the ``pursued'' DoF by user $k$, that is the number of symbols it wishes to transmit.
To aid intuition, we note that achievable DoF can be perceived as the number of signal space dimensions that are {\it free} of interference. 
Prior to transmitting, user $k\in\mathcal{K}$ linearly precodes its symbol vector to obtain ${\bf s}_k\eqdef{\bf V}_k{\bf x}_k$, where ${\bf V}_k\in\mathbb{C}^{M_{\text{t}}\times d}$ denotes the precoding matrix
 whose $d$ columns are linearly independent. 
We consider  signal vectors with an expected power constraint $E\left\{\|{\bf s}_k\|^2\right\}\le P$, for all $k\in\mathcal{K}$.
The downconverted and pulse matched received signal at receiver $k$  is given by
\begin{align}
{\bf y}_k&\eqdef{\bf H}_{k,k}{\bf s}_k+\sum_{l=1,l\ne k}^K{\bf H}_{k,l}{\bf s}_l+{\bf w}_k\nonumber\\
&={\bf H}_{k,k}{\bf V}_k{\bf x}_k+\sum_{l=1,l\ne k}^K{\bf H}_{k,l}{\bf V}_l{\bf x}_l+{\bf w}_k,
\end{align}
where ${\bf H}_{i,j}\in\mathbb{C}^{M_{\text{r}}\times  M_{\text{t}}}$ represents the channel ``processing'' between the $j$th user and the $i$th receiver and ${\bf w}_k\in\mathbb{C}^{M_{\text{r}}\times  1}$ denotes the zero-mean complex additive white Gaussian noise vector with covariance matrix $\sigma_k^2{\bf I}_{M_{\text{r}}}$, where $i,j,k\in\mathcal{K}$. 
Each receiver $k\in \mathcal{K}$, linearly processes the received signal to obtain 
\begin{equation}
{\bf U}^H_k{\bf y}_k={\bf U}^H_k{\bf H}_{k,k}{\bf V}_k{\bf x}_k+{\bf U}^H_k\sum_{l=1,l\ne k}^K{\bf H}_{k,l}{\bf V}_l{\bf x}_l+{\bf U}^H_k{\bf w}_k,
\end{equation}
where ${\bf U}_k\in\mathbb{C}^{M_{\text{r}}\times  d}$ is the corresponding linear zero-forcing filter
 with $d$ linearly independent columns. 
Hence, $\text{span}\left({\bf U}^H_k{\bf H}_{k,k}{\bf V}_k\right)$ constitutes the useful signal space in which receiver $k\in \mathcal{K}$  expects to observe the symbols transmitted by user $k$, while $\text{span}\left(\left\{{\bf U}^H_k{\bf H}_{k,l}{\bf V}_l\right\}_{l=1,l\ne k}^{K}\right)$ is the space where all interference is observed. We denote by $\{{\bf X}_l\}_{l=1,l\ne k}^K$ the horizontal concatenation of matrices ${\bf X}_1,\ldots, {\bf X}_{k-1},{\bf X}_{k+1},\ldots{\bf X}_K$.
We denote this $K$-user MIMO interference channel as an $(M_{\text{r}}\times M_{\text{t}},d)^K$  system, in the same manner as in \cite{Yetis}, where all signal spaces span all available dimensions.
For all the cases considered we assume $d\le\min(M_{\text{t}},M_{\text{r}})$.

For practical reasons one might consider ${\bf V}_k^H{\bf V}_k=\frac{P}{d}{\bf I}_d$, for all $k\in\mathcal{K}$. This might be a setting where each column of ${\bf V}_k$ represents a beamforming (or signature) vector assigned to a user in a group  (or a cell) of $d$ users and enforces orthogonality among user signal subspaces. 
Accordingly, we may as well assume that the columns of each zero-forcing filter ${\bf U}_k$, for all receivers $k\in\mathcal{K}$, form an $d$-dimensional orthonormal basis, if practical interest requires such a construction.

\begin{rem}
 We note that for simplicity we made the assumption that the number of transmit antenna elements are the same at all transmitters or receivers and all users wish to transmit $d>0$ symbols. 
However, the results that follow can be easily carried to the case where transmitters and receivers might not have the same  number of antennas.
\end{rem}

\section{Interference Alignment as a Rank Constrained Rank Minimization}
In this section we show that for each user $k\in\mathcal{K}$, the maximum achievable DoF can be put in the form of an RCRM problem.
We suggest a new optimization framework for the Interference Alignment problem instead of the one for {\it perfect} IA as presented in \cite{Gomadam}. 
The optimization aims for the maximum interference suppression possible, when the signal space of each user is required to span exactly $d$ spatial dimensions worth of communication.
The perfect IA conditions being stricter require that interference signals are nulled out using linear processing of the received signal at each receiver, when the useful signal space spans $d$ dimensions. 
To generalize the perfect IA framework, we use ranks of matrices to account for the number of useful signaling and interference  dimensions. 
Interestingly, we show that the RCRM formulation is equivalent to the sum DoF maximization in an $(M_{\text{r}}\times M_{\text{t}}, d)^K$ system, where each useful signal space spans $d$ dimensions.
Then, we use this framework to develop a new approximation algorithm and compare its tightness to existing interference leakage minimization approaches.

We begin by stating the perfect IA requirements. In an $(M_{\text{r}}\times M_{\text{t}}, d)^K$ system and for all $ k\in\mathcal{K}$ perfect IA requires that
\begin{align}
{\bf U}_k^H{\bf H}_{k,l}{\bf V}_l&={\bf 0}_{d\times d},\;\;\; \forall l\in\mathcal{K}\backslash k,\label{req1}\\
\rank\left({\bf U}_k^H{\bf H}_{k,k}{\bf V}_k\right)&=d\label{req2},
\end{align}
where (\ref{req1}) enforces all interference spaces to have zero dimensions posterior to zero-forcing and (\ref{req2}) enforces the useful signal to span all $d$ dimensions. 

\begin{rem}
Observe that (\ref{req1}) is a set of bilinear equations in the unknown precoding and zero-forcing filters. 
Recently, a feasibility question has been raised as to whether a system admits perfect interference alignment or not. 
When the channel coefficients are selected randomly and independently, \cite{Yetis} claimed that proper systems
should have a perfect alignment solution almost surely. 
An  $(M_{\text{r}}\times M_{\text{t}}, d)^K$ system is called {\it proper}, i.e. perfect IA is expected to be feasible almost surely, when the number of variables is less than or equal to the number of equations, and this is equivalent to $M_{\text{r}}+M_{\text{t}}-d(K+1)\ge0$.  
\end{rem}

Interestingly, if we look at the problem of finding the precoding and zero-forcing matrices
when the channel coefficients are arbitrary and given, the problem is computationally intractable.
In particular, Razaviyayn \textit{et al} recently established that for an arbitrary $K$-user MIMO interference channel, checking the achievability of a certain DoF tuple $\{d_1,\ldots,d_K\}$  is NP-hard when each user and transmitter has more than $2$ antennas \cite{Luo}.
Therefore, solving the perfect IA set of bilinear equations in a single-shot manner cannot be performed efficiently, in the general case. 
Additionally, scenaria of multiple symbol extensions for the MIMO case, or multi-cell systems as the ones presented in \cite{Suh}, lack a similar definition to a proper system, however the sum DoF maximization is still an achievement one should be aiming for.
Hence, it is interesting to obtain a more general framework for the IA problem that not only captures the notion of what are the maximum DoF conditions, but also quantifies what one should be optimizing to achieve this DoF.
Of course, a hard optimization problem might be equivalently or even harder than solving a set of bilinear equations.
However, generalizing the formulation will provide helpful cues for good approximation schemes and better understanding of the problem.

We continue with rewriting (\ref{req1})
\begin{equation}
\begin{split}
{\bf U}_k^H{\bf H}_{k,l}{\bf V}_l&={\bf 0}_{d\times d},\;\;\; \forall l\in\mathcal{K}\backslash k\\
\Leftrightarrow \left[\left\{{\bf U}_k^H{\bf H}_{k,l}{\bf V}_l\right\}_{l=1,l\ne k}^K\right]&=\left[{\bf 0}_{d\times d}\ldots{\bf 0}_{d\times d}\right]\\
\Leftrightarrow {\bf U}_k^H\left[\left\{{\bf H}_{k,l}{\bf V}_l\right\}_{l=1,l\ne k}^K\right]&={\bf 0}_{d\times (K-1)d}, \nonumber
\end{split}
\end{equation}
and defining the {\it signal} and {\it interference} matrices for all $k\in\mathcal{K}$
{\small
\begin{align}
{\bf S}_k\left({\bf U}_k,{\bf V}_k\right)&\eqdef{\bf U}_k^H{\bf H}_{k,k}{\bf V}_k\in\mathbb{C}^{d\times d},\\
{\bf J}_k\left({\bf U}_k,\left\{{\bf V}_l\right\}_{l=1}^K\right)&\eqdef{\bf U}_k^H\left[\left\{{\bf H}_{k,l}{\bf V}_l\right\}_{l=1,l\ne k}^K\right]\in\mathbb{C}^{d\times (K-1)d}.
\end{align}
}For ease of notation we refer to ${\bf S}_k\left({\bf U}_k,{\bf V}_k\right)$ and ${\bf J}_k\left({\bf U}_k,\left\{{\bf V}_l\right\}_{l=1}^K\right)$ as ${\bf S}_k$ and ${\bf J}_k$, respectively.
The space spanned by the columns of ${\bf S}_k$ is the space in which the $k$th receiver expects to observe the transmitted signal ${\bf x}_k$. Accordingly, the space spanned by the columns of ${\bf J}_k$ accounts for the interference space at receiver $k\in\mathcal{K}$.
Now, we restate (\ref{req1}) and (\ref{req2}) in terms of ranks 
\begin{align}
\rank\left({\bf J}_k\right)&=0,\label{myreq1}\\
\rank\left({\bf S}_k\right)&=d,\label{myreq2}
\end{align}
 for all $k\in \mathcal{K}$.

To motivate the rank notation we use the following lemma to characterizes the spatial degrees of freedom for a given user $k$  of an $(M_{\text{r}}\times M_{\text{t}},d)^K$ static flat-fading MIMO interference system.
\begin{lem}
Let $\left\{{\bf V}_l\right\}_{l=1}^K$ be a given set of precoding filters and ${\bf U}_k$ be a given zero-forcing filter employed by user $k\in\mathcal{K}$.
Then, the achievable spatial DoF by user $k$ for these sets is
\begin{equation}
{d}_k \eqdef\left[\rank\left({\bf S}_k\right)-\rank\left({\bf J}_k\right)\right]^+.
\end{equation}
\end{lem}
{\bf Proof}:
We have for $k\in\mathcal{K}$
{\small
\begin{align}
{d}_k&=\frac{I\left({\bf U}{\bf s}_k;{\bf U}{\bf y}_k\right)}{\log(P)}\nonumber\\
& = \lim_{P\rightarrow\infty}\frac{\log\det\left( {\bf I}_d+ \left({\bf J}_k {\bf J}^H_k+\sigma^2{\bf I}_d\right)^{-1}{\bf S}_k {\bf S}^H_k\right)}{\log(P)}\nonumber\\
& = \left[\lim_{P\rightarrow\infty}\frac{\log\det\left( \left({\bf J}_k {\bf J}^H_k+\sigma^2{\bf I}_d\right)^{-1}{\bf S}_k {\bf S}^H_k\right)}{\log(P)}\right]^+\nonumber\\
&=\left[\lim_{P\rightarrow\infty}\frac{\log\det \left({\bf S}_k {\bf S}^H_k\right)-\log\det \left({\bf J}_k {\bf J}^H_k+\sigma^2{\bf I}_d\right)}{\log(P)}\right]^+\nonumber\\
&=\Biggl[\lim_{P\rightarrow\infty}\frac{\sum_{i=1}^{\rank\left({\bf S}_k\right)}\log\left(\lambda_i\left({\bf S}_k {\bf S}^H_k\right)\right)}{\log(P)}\\
&-\frac{\sum_{i=1}^{\rank\left({\bf J}_k\right)}\log\left(\lambda_i\left({\bf J}_k {\bf J}^H_k\right)+\sigma^2\right)}{\log(P)}\Biggr]^+\nonumber\\
&=\left[\lim_{P\rightarrow\infty}\frac{\sum_{i=1}^{\rank\left({\bf S}_k\right)}\log(P)-\sum_{i=1}^{\rank\left({\bf J}_k\right)}\log(P)}{\log(P)}\right]^+\nonumber\\
&=\left[\rank\left({\bf S}_k\right)-\rank\left({\bf J}_k\right)\right]^+,
\end{align}
}
where $\left[a\right]^+=\left\{\begin{array}{cl}a&,a>0\\0&,a\le0\end{array}\right.$ and $\lambda_i({\bf A})$ is the $i$th largest eigenvalue of ${\bf A}$.
The above result is possible due to the eigenvalues of the ${\bf S}_k{\bf S}_k^H$ and ${\bf J}_k{\bf J}_k^H$ scaling linearly with $P$ for given precoding, zero-forcing, and channel matrices.
\hfill$\Box$\\

Lemma 1 implies that to maximize the per user DoF we have to design transmit and receive strategies in a sophisticated way, such that this difference of ranks is maximized. 
More precisely, for an $(M_{\text{r}}\times M_{\text{t}},d)^K$ system where each user aims for $d$ DoF, the following
 set of $K$ ``parallel'' rank constrained rank minimization problems has to be solved
\begin{align}
&\hspace{-0.7cm}\underset{\left\{{\bf V}_l\right\}_{l=1,l\ne k}^K,{\bf U}_k}{{\min}}\hspace{-0.5cm}\rank\left({\bf J}_k\right)\label{minreq1}\\
\text{s.t.: }&\rank\left({\bf S}_k\right)=d.\label{minreq2}
\end{align}
It is obvious, that when perfect IA is possible (\ref{minreq1}) and (\ref{minreq2}) will find it, or else the best possible solution will be obtained for user $k\in\mathcal{K}$, which will be equal to 
\begin{align}
&\hspace{-0.7cm}\underset{\left\{{\bf V}_l\right\}_{l=1,l\ne k}^K,{\bf U}_k}{\max}\left(\rank({\bf S}_k)-\rank\left({\bf J}_k\right)\right)\\
\text{s.t.: }&\rank\left({\bf S}_k\right)=d\nonumber.
\end{align}

\begin{rem}
As noted before, for the constant $K$-user MIMO interference channel it is known that the maximum per user achievable DoF using linear precoding and zeroforcing schemes is $d^*=\frac{M_{\text{t}}+M_{\text{r}}}{K+1}$, almost surely \cite{Yetis}, hence aiming for more than $d^*$ seems unmotivated. 
In that regard, when aiming for $d\le d^*$, we should expect that optimally the rank of each interference can go down to zero, as perfect IA requires. 
Hence, why use the rank formulation? 
As we see in the following, the fact that the rank formulation completely captures the notion of per user DoF assists us in tightly approximating this objective. 
The nature of our approximation favors solutions where the interference has low rank (i.e., users get higher DoF), instead of low energy (i.e., small interference leakage). 
This is of particular interest when perfect IA is feasible but hard to obtain, or for interference channels where the notion of proper is not clearly defined, such as cellular networks.
\end{rem}

Apparently, it is not trivial to solve in parallel a set of the $K$ optimization problems in (\ref{minreq1}) and (\ref{minreq2}). 
Alternatively, we can maximize the sum-DoF of all $K$ users through the following RCRM.

  \begin{equation}
\boxed{
\begin{split}
\mathcal{P}:&\\
&\underset{\text{\scriptsize{$\begin{array}{c}\left\{{\bf V}_l\right\}_{l=1}^K\\\left\{{\bf U}_l\right\}_{l=1}^K\end{array}$}}} {\min}\sum_{k=1}^K\rank\left({\bf J}_k\right)\\
\text{s.t.: }&\rank\left({\bf S}_k\right)=d,\;\;\forall k\in\mathcal{K}.\nonumber
\end{split}
}
  \end{equation}

\begin{theo}
A solution to $\mathcal{P}$ maximizes the sum of spatial degrees of freedom for an $(M_{\text{r}}\times M_{\text{t}},d)^K$ static flat-fading MIMO interference channel, when every signal space spans $d$ dimensions.
\end{theo}
{\bf Proof:} 
For all selections of ${\bf V}_1,\ldots,{\bf V}_K\in\mathbb{C}^{M_{\text{t}}\times  d}$ and ${\bf U}_1,\ldots,{\bf U}_K\in\mathbb{C}^{M_{\text{r}}\times  d}$ satisfying the constraints of $\mathcal{P}$ we have that
{\small
\begin{equation}
\begin{split}
\sum_{k=1}^K{d}_k & = \sum_{k=1}^K\left[\rank\left({\bf S}_k\right)-\rank\left({\bf J}_k\right)\right]^+= Kd-\sum_{k=1}^K\rank\left({\bf J}_k\right)\\
\Leftrightarrow&\underset{\text{\scriptsize{$\begin{array}{c}\left\{{\bf V}_l\right\}_{l=1}^K\\\left\{{\bf U}_l\right\}_{l=1}^K\end{array}$}}}{\arg\max}\sum_{k=1}^K {d}_k=\underset{\text{\scriptsize{$\begin{array}{c}\left\{{\bf V}_l\right\}_{l=1}^K\\\left\{{\bf U}_l\right\}_{l=1}^K\end{array}$}}}{\arg\max}\left\{Kd-\sum_{k=1}^K\rank\left({\bf J}_k\right)\right\}\\
\Leftrightarrow&\underset{\text{\scriptsize{$\begin{array}{c}\left\{{\bf V}_l\right\}_{l=1}^K\\\left\{{\bf U}_l\right\}_{l=1}^K\end{array}$}}}{\arg\max}\sum_{k=1}^K {d}_k=\underset{\text{\scriptsize{$\begin{array}{c}\left\{{\bf V}_l\right\}_{l=1}^K\\\left\{{\bf U}_l\right\}_{l=1}^K\end{array}$}}}{\arg\min}\sum_{k=1}^K\rank\left({\bf J}_k\right)\nonumber.
\end{split}
\end{equation}
}
\hfill$\Box$

The orthogonality constraints for the precoding and zero-forcing matrices were ommited in the RCRM, since we can always linearly transform them so that the columns of each of these matrices are mutually orthogonal. 
Namely, we can rewrite the precoding matrices as ${\bf V}_k={\bf Q}^{(v)}_k{\bf R}^{(v)}_k$, where ${\bf Q}^{(v)}_k\in\mathbb{C}^{M_{\text{t}}\times  d}$ is an orthonormal basis for the column space of ${\bf V}_k$ and ${\bf R}^{(v)}_k\in\mathbb{C}^{d\times d}$ is the matrix of coefficients participating in the linear combinations yielding the columns of ${\bf V}_k$. Then we may use $\sqrt{\frac{P}{d}}{\bf Q}^{(v)}_k$ as the precoding matrix. 
Accordingly, we use orthonormal matrices ${\bf Q}^{(u)}_k$ constructed by decomposing ${\bf U}_k$ to ${\bf Q}^{(u)}_k{\bf R}^{(u)}_k$, where ${\bf Q}^{(u)}_k\in\mathbb{C}^{M_{\text{r}}\times  d}$ and ${\bf Q}^{(u)}_k\in\mathbb{C}^{d\times d}$. Observe that 
$\text{span}\left({\bf Q}^{(v)}_k\right)=\text{span}\left({\bf V}_k\right)\text{ and }\text{span}\left({\bf Q}^{(u)}_k\right)=\text{span}\left({\bf U}_k\right)$
for all $k\in\mathcal{K}$.
Moreover, the ranks of the interference and signal matrices are oblivious to full-rank linear transformations
{\footnotesize
\begin{align}
&\rank\left({\bf J}_k\right)\nonumber \\
&\hspace{-0.1cm}= \rank\left(\hspace{-0.1cm}\left({\bf R}^{(u)}_k\right)^H\left({\bf Q}^{(u)}_k\right)^H\left[\left\{{\bf H}_{k,l}{\bf V}_l\right\}_{l=1,l\ne k}^K\right]\hspace{-0.1cm}\right)\nonumber\\
&\hspace{-0.1cm}= \rank\left(\hspace{-0.1cm}\left({\bf Q}^{(u)}_k\right)^H\hspace{-0.1cm}\left[\left\{{\bf H}_{k,l}{\bf Q}^{(v)}_l\right\}_{l=1,l\ne k}^K\right]\text{blkdiag}\hspace{-0.1cm}\left(\hspace{-0.1cm}\left\{{\bf R}^{(v)}_l\right\}_{l=1,l\ne k}^K\right)\hspace{-0.1cm}\right)\nonumber\\
&\hspace{-0.1cm}= \rank\left(\left({\bf Q}^{(u)}_k\right)^H\left[\left\{{\bf H}_{k,l}{\bf Q}^{(v)}_l\right\}_{l=1,l\ne k}^K\right]\right),\label{QR}
\end{align}
}where $\text{blkdiag}\left({\bf A}_1,\ldots,{\bf A}_n\right)$ denotes the block diagonal matrix that has  as $i$th diagonal block the matrix ${\bf A}_i$, the above equalities hold due to the fact that $\rank\left({\bf R}_l^{(u)}\right)=\rank\left({\bf R}_l^{(v)}\right)=d$. Furthermore, we have
\begin{equation}
\begin{split}
\hspace{-0.12cm}\rank\left({\bf S}_k\right) 
&= \rank\left(\left({\bf R}^{(u)}_k\right)^H\left({\bf Q}^{(u)}_k\right)^H{\bf H}_{k,k}{\bf Q}^{(v)}_k{\bf R}^{(v)}_k\right)\\
&= \rank\left(\left({\bf Q}^{(u)}_k\right)^H{\bf H}_{k,k}{\bf Q}^{(v)}_k\right)=d.\nonumber
\end{split}
\end{equation}
Hence, orthogonalization is always possible when $d\le\min(M_{\text{t}},M_{\text{r}})$.

\begin{rem}
Observe that we can generalize $\mathcal{P}$ to the case where  user $k$ is equipped with $M_{\text{t},k}$ antennas,  receiver $k$ with $M_{\text{r},k}$ antennas, and the $k$ user's signal space spans $d_k$ dimensions, $k\in\mathcal{K}$. 
This generalized version of $\mathcal{P}$ can be used to decide the achievability of any DoF tuple $\{d_1,d_2,\ldots,d_K\}$: if this tuple is achievable the cost function of $\mathcal{P}$ will be zero.
However, this is an NP-hard problem for the case of $M_{\text{t},k},M_{\text{r},k}>2$ \cite{Luo}. 
Therefore, the generalized version of $\mathcal{P}$ has to be at least as hard as determining the achievability of a DoF tuple. 
\end{rem}

To conclude, we have established that the minimization of the sum of the interference dimensions under full rank signal space constraints is equivalent to maximizing the sum DoF of a static flat-fading MIMO interference channel. 
There exist various instances and regimes of difficulty for this RCRM problem. 
There are tractable regimes where one randomly selects the precoding or zero-forcing matrices matrices and constructs the zero-forcing or precoding matrices with columns that exactly fit in the null space of the interference or reciprocal interference matrices. 
This is possible when either $d\le M_{\text{r}}-(K-1)d$, or $d\le M_{\text{t}}-(K-1)d$ holds. 
Moreover, when the channel matrices are diagonal, the symbol extension method presented in \cite{Cadambe1} creates instances of the RCRM problem that can be efficiently solved asymptotically.
However, in the general case such solutions are NP-hard to obtain  and the RCRM problem cannot be efficiently solved.
In the next section we provide a heuristic that approximates $\mathcal{P}$.

\section{A Nuclear Norm Heuristic}

In the previous section we establish that maximizing the sum DoF of a $K$-user MIMO interference channel is equivalent to solving an RCRM, where the precoding and zero-forcing matrices are the optimization variables.
To approach this highly nonconvex and intractable problem we use convex relaxations for the cost function and constraints of $\mathcal{P}$. 

We begin by obtaining the tightest convex approximation to the cost function of $\mathcal{P}$. 
We have
{\small
\begin{align}
 \overline{\text{conv}}\left(\sum_{k=1}^K\rank\left({\bf J}_k\right)\right)&= 
\overline{\text{conv}}\left(\rank\left(\text{blkdiag}\left(
{\bf J}_1,\ldots, {\bf J}_K \right)\right)\right)\nonumber\\
&=\frac{1}{\mu}\left\|
\text{blkdiag}\left(
{\bf J}_1,\ldots, {\bf J}_K \right)
\right\|_*\label{nuclear}\\
&= \frac{1}{\mu}\sum_{k=1}^K\left\|{\bf J}_k\right\|_*=\frac{1}{\mu}\sum_{k=1}^K\sum_{i=1}^d\sigma_i\left({\bf J}_k\right), \nonumber
\end{align}}where $\overline{\text{conv}}(f)$ denotes the convex envelope of a function $f$, $\|{\bf A}\|_*=\sum_{i=1}^{\rank({\bf A})}\sigma_i({\bf A})$ is the nuclear norm of a matrix ${\bf A}$ which accounts for the sum (i.e., the $\ell_1$-norm) of the singular values of ${\bf A}$, and $\sigma_i\left({\bf A}\right)$ is the $i$th largest singular value of ${\bf A}$. 
This normalized sum of nuclear norms in (\ref{nuclear}) is the convex envelope of the sum of interference ranks when the maximum singular value of the interference matrices is upper bounded by $\mu>0$ \cite{Recht}.
More formally,  this is possible when we operate on the following sets of interference matrices
$\bigl\{{\bf J}_k\in\mathbb{C}^{d\times (K-1)d},\;\forall k\in\mathcal{K};
\max_{k\in\mathcal{K}}\sigma_1\left({\bf J}_k\right)\le \mu\bigr\}$.

Before we proceed with approximating nonconvex rank constraints by a convex feasible set, we provide some insights on the algorithms presented in \cite{Gomadam} and \cite{Peters1}. These algorithms aim to minimize, in alternating optimization fashion, the total interference leakage at each receiver, a metric defined as
\begin{equation}
\sum_{k=1}^K\tr\left\{{\bf U}_k^H{\bf Q}_k{\bf U}_k\right\}\label{leak},
\end{equation}
where
\begin{equation}
{\bf Q}_k=\sum_{l=1,l\ne k}^K \frac{P}{d}{\bf H}_{k.l}{\bf V}_l{\bf V}_l^H{\bf H}_{k,l}^H,\label{QI}
\end{equation}
subject to orthogonality constraints of the columns of the precoding and zero-forcing matrices, i.e., ${\bf V}_k^H{\bf V}_k=\frac{P}{d}{\bf I}_d$ and ${\bf U}_k^H{\bf U}_k={\bf I}_d$, for all $k\in\mathcal{K}$.
At each step of the optimization, either the precoding or zero-forcing matrices are fixed, and minimization is performed over the free variables.
Observe, that if we plug (\ref{QI}) in (\ref{leak}) we get
{\small
\begin{equation}
\begin{split}
&\sum_{k=1}^K\tr\left\{{\bf U}_k^H\left(\sum_{l=1,l\ne k}^K \frac{P}{d}{\bf H}_{k,l}{\bf V}_l{\bf V}_l^H{\bf H}_{k,l}^H\right){\bf U}_k\right\}\\
&=\frac{P}{d}\sum_{k=1}^K\tr\left\{{\bf U}_k^H\left[\left\{{\bf H}_{k,l}{\bf V}_l\right\}_{l=1,l\ne k}^K\right]\left[\left\{{\bf H}_{k,l}{\bf V}_l\right\}_{l=1,l\ne k}^K\right]^H
{\bf U}_k\right\}\\
&=\frac{P}{d}\sum_{k=1}^K\tr\left\{{\bf J}_k{\bf J}^H_k\right\}=\frac{P}{d}\sum_{k=1}^K\left\|{\bf J}_k\right\|_F^2\\
&=\frac{P}{d}\sum_{k=1}^K\sum_{i=1}^{d}\sigma_i^2\left({\bf J}_k\right),
\end{split}
\end{equation}
}where $\|{\bf A}\|_F$ is the Frobenius norm of ${\bf A}$ and the constant $\frac{P}{d}$ can be dropped in the minimization. 
Therefore, the interference leakage metric is the $\ell_2$ norm of the singular values of all interference matrices, i.e. the sum of ``interference energy'' at all receivers.

Experimentally, it has been observed that the alternating leakage minimization performs well and yields perfect IA solutions for various instances of proper systems. 
However, when it does not yield perfect IA for systems that are proper, it outputs precoding and zero-forcing matrices that result to ``low'' energy  interference, i.e., interference singular values with small $\ell_2$-norm.
For such solutions interference may be weak, nonetheless its span is not confined to  low dimensional subspaces.
Potentially, it spans more dimensions than the sparsest possible solution.
Moreover, there are known configurations for which it cannot output solutions that give full rank signal spaces. These are cases of channel matrices with specific structure, such as symbol extended channels.
There are also some interesting setups, such as cellular interference channels, where the leakage minimization approaches have not yet been generalized to and sum-DoF maximization is again a well motivated objective to pursue.

Here, we claim that minimizing the $\ell_1$-norm of the interference singular values is intuitively expected to output low-rank solutions, as it is provably the case for affine constrained, rank minimization problems \cite{Recht}, \cite{Candes}. 
Low-rank interference is what we were aiming for in the first place, so to maximize the sum-DoF.
However, we should note that our problem formulation, although a rank minimization itself, does not fit exactly in the affine constrained, rank minimization framework of  \cite{Candes} and exact solution guarantees cannot be used for our case in a straightforward manner.

We continue our approximation by obtaining constraints defining a convex and tractable feasible set. 
We approximate the constraint, $\rank\left({\bf S}_k\right)=d,\;\;\forall k\in\mathcal{K}$ with 
\begin{align}
&\lambda_{\text{min}}\left({\bf S}_k\right)\ge\epsilon,\label{constr1}\\
&{\bf S}_k\succeq {\bf 0}_{d\times d},\label{constr2}
\end{align}
where $\lambda_{\text{min}}\left({\bf S}_k\right)$ is the minimum eigenvalue of ${\bf S}_k$, $\epsilon>0$, and ${\bf S}_k\succeq {\bf 0}_{d\times d}$ denotes that matrix ${\bf S}_k$ is hermitian positive semidefinite; that is for all $k\in\mathcal{K}$ we have $\left\{{\bf S}_k\in\mathbb{C}^{d\times d}; {\bf S}_k={\bf S}_k^H \text{ and }{\bf z}^H{\bf S}_k{\bf z}\ge0,\; \forall {\bf z}\in\mathbb{C}^d\right\}$. 
We note that the minimum eigenvalue constraint on positive semidefinite matrices serves as a tractable constraint yielding a convex feasible solution set. 
This relaxation might seem stricter than the rank constraints, however, this is not the case (considering the initial cost function of the sum of ranks) when we replace the closed set defined by $\lambda_{\text{min}}\left({\bf S}_k\right)\ge\epsilon$ with the open set defined by  $\lambda_{\text{min}}\left({\bf S}_k\right)>0$. 
Although, the former set is always a subset of the latter, as $\epsilon$ gets smaller, the two sets will asymptotically overlap.
We proceed by proving that for any feasible solution of the RCRM problem there exists (at least) one solution that gives positive definite useful space matrices and preserves all interference and signal space rank properties.
\begin{lem}
Let $\{{\bf V}_l\}_{l=1}^K$ and $\{{\bf U}_l\}_{l=1}^K$ be any feasible pair of sets of precoding and zero-forcing matrices so that $\rank({\bf J}_k)=\rho_k$ and $\rank\left({\bf S}_k\right)=d$, for any $k\in\mathcal{K}$. 
Then, for any such feasible point of $\mathcal{P}$ there exists a feasible pair of sets for (\ref{constr1}) and (\ref{constr2}), that is
\begin{align}
\{\hat{{\bf V}}_k\}_{k=1}^K&=\{{\bf V}_k{\bf V}^H_k{\bf H}^H_{k,k}{\bf U}_k\}_{k=1}^K\\
\text{and }\{\hat{{\bf U}}_k\}_{k=1}^K&=\{{\bf U}_k{\bf U}^H_k{\bf H}_{k,k}{\bf V}_k\}_{k=1}^K,
\end{align} 
such that
{\small 
\begin{align}
&\rank\left({\bf J}_k\left({\bf U}_k,\left\{\hat{\bf V}_l\right\}_{l=1,l\ne k}^K\right)\right)\nonumber\\
&=\rank\left({\bf J}_k\left(\hat{\bf U}_k,\left\{{\bf V}_l\right\}_{l=1,l\ne k}^K\right)\right)=\rho_k\label{rank1}\\
&\rank\left({\bf S}_k\left({\bf U}_k,\hat{\bf V}_k\right)\right)\nonumber\\
&=\rank\left({\bf S}_k\left(\hat{\bf U}_k,{\bf V}_k\right)\right)=d,\label{rank2}\\
&\lambda_{\min}\left({\bf S}_k\left({\bf U}_k,\hat{\bf V}_k\right)\right)>0,\text{ and }{\bf S}_k\left({\bf U}_k,\hat{\bf V}_k\right)\succ0\label{psd1},\\
\text{ and }&\lambda_{\min}\left({\bf S}_k\left(\hat{\bf U}_k,{\bf V}_k\right)\right)>0,\text{ and }{\bf S}_k\left(\hat{\bf U}_k,{\bf V}_k\right)\succ0\label{psd2},
\end{align}
}for any $k\in\mathcal{K}$.
\end{lem}
{\bf Proof}:
Observe that 
{\small
\begin{equation}
\begin{split}
&\rank\left( {\bf J}_k\right)\\
&=\rank\left( {\bf U}^H_k\left[\left\{{\bf H}_{k,l}{\bf V}_l\right\}_{l=1,l\ne k}^K\right]\right)\\
&=\rank\left( {\bf A}_k{\bf U}^H_k\left[\left\{{\bf H}_{k,l}{\bf V}_l\right\}_{l=1,l\ne k}^K\right]\text{blkdiag}\left({\bf A}_1,\ldots, {\bf A}_{k-1}\right)\right)\\
&=\rho_k,
\end{split}
\end{equation}
}for any set of full-rank matrices ${\bf A}_1,\ldots,{\bf A}_k\in\mathbb{R}^{d\times d}$. 
Using this argument, (\ref{rank1}) is true since 
${\bf V}^H_k{\bf H}^H_{k,k}{\bf U}_k$ and its hermitian are square, full-rank matrices for all $k\in\mathcal{K}$.
Moreover, (\ref{rank2}) is true due to the same argument, that is,
\begin{align}
&\rank\left({\bf U}^H_k{\bf H}_{k,k}{\bf V}_k\right)\\
&=\rank\left({\bf U}^H_k{\bf H}_{k,k}\hat{\bf V}_k\right)\\
&=\rank\left({\bf U}^H_k{\bf H}_{k,k}{\bf V}_k{\bf V}^H_k{\bf H}^H_{k,k}{\bf U}_k\right)\\
&=\rank\left(\hat{\bf U}^H_k{\bf H}_{k,k}{\bf V}_k\right)\\
&=\rank\left({\bf V}^H_k{\bf H}^H_{k,k}{\bf U}_k{\bf U}^H_k{\bf H}_{k,k}{\bf V}_k\right)=d.
\end{align}
Moreover, it is straightforward that
\begin{align}
&\lambda_{\text{min}}\left({\bf U}^H_{k}{\bf H}_{k,k}{\bf V}_k{\bf V}^H_k{\bf H}^H_{k,k}{\bf U}_k\right)\nonumber\\
&=\min_{\|{\bf x}\|_2=1}\left\|{\bf V}^H_k{\bf H}^H_{k,k}{\bf U}_k{\bf x}\right\|^2_2>0,\\
&\lambda_{\text{min}}\left({\bf V}^H_k{\bf H}^H_{k,k}{\bf U}_k{\bf U}^H_k{\bf H}_{k,k}{\bf V}_k\right)\nonumber\\
&=\min_{\|{\bf x}\|_2=1}\left\|{\bf U}_k{\bf H}_{k,k}{\bf V}_k{\bf x}\right\|^2_2>0,
\end{align}
thus, (\ref{psd1}) and (\ref{psd2}) are true. 
\hfill$\Box$

Having provided relaxations for both the cost function and constraints of $\mathcal{P}$, we are stating the steps of our approximation algorithm.
First, we decide to arbitrarily select  the zero-forcing matrices. 
Then, we solve the following convex optimization problem
 \begin{equation}
\boxed{
\begin{split}
\mathcal{A}_{{\bf V}}\left(\left\{{\bf U}_l\right\}_{l=1}^K\right):&\\
&\underset{\left\{{\bf V}_l\right\}_{l=1}^K}{\min}\sum_{k=1}^K\left\|{\bf J}_k\right\|_*\\
\text{s.t.: }&\lambda_{\min}\left({\bf S}_k\right)\ge\epsilon,\\
&{\bf S}_k\succeq {\bf 0}_{d\times d}, \; \forall k\in\mathcal{K}.\nonumber
\end{split}
}
\end{equation}
use its solution as an input to $\mathcal{A}_{{\bf U}}\left(\left\{{\bf V}_l\right\}_{l=1}^K\right)$, and then feed the solutions of this optimization  back to $\mathcal{A}_{{\bf V}}\left(\left\{{\bf U}_l\right\}_{l=1}^K\right)$. We continue by iterating this process.
This procedure is stated below as algorithm $\mathcal{A}(n)$, where $n$ is the number of iterations. 
\begin{center}
\begin{tabular}{|ll|}
\hline
$\mathcal{A}(n)$:&\\
\hline
\texttt{1}:&\texttt{initialize} $\left\{{\bf U}_l\right\}_{l=1}^K$\\
\texttt{2}:&\texttt{for $n$ iterations}\\
\texttt{3}:&\hspace{0.5cm}$\left\{{\bf V}_l\right\}_{l=1}^K\leftarrow\mathcal{A}_{{\bf V}}\left(\left\{{\bf U}_l\right\}_{l=1}^K\right)$\\
\texttt{4}:&\hspace{0.5cm}$\left\{{\bf U}_l\right\}_{l=1}^K\leftarrow\mathcal{A}_{{\bf U}}\left(\left\{{\bf V}_l\right\}_{l=1}^K\right)$\\
\texttt{5}:&\texttt{orthogonalize}  $\left\{{\bf V}_l\right\}_{l=1}^K$, $\left\{{\bf U}_l\right\}_{l=1}^K$\\
\hline
\end{tabular}
\end{center}
Observe that this iterative procedure is bound to converge to a local optimum. 
In the simulations section we provide quantitative results for the performance of our proposed approximation and observe that it indeed favors matrices that yield low-rank interference solutions for our problem.

\begin{rem}
We would like to note a key difference between our iterative approach and the approaches  of \cite{Gomadam} and \cite{Peters1}.
For channels that do not have the block diagonal structure, (\ref{req2}) almost surely holds when using the alternating leakage minimization approach due to the precoding and zero-forcing orthogonality constraints.
However, this probabilistic argument does no longer hold, at least for the leakage minimization scheme, when considering channels with block diagonal structure.
Hence, zero interference leakage may be obtained using alternating leakage minimization, but at the same time the signal spaces may be confined to less than $d$ dimensions not obeying (\ref{req2}) \cite{Gomadam2}.
Interestingly, our approach is robust with respect to channel structures since it explicitly involves a positivity constraint on the minimum eigenvalue of the signal space matrix, that is (\ref{req2}) will always hold for {\it any} set of channel matrices.
\end{rem}

To conclude this section, we have introduced an alternating nuclear norm minimization scheme to approximate the sum DoF maximization in the $k$-user MIMO interference channel.
Our approximation is motivated by our RCRM formulation for the sum DoF maximization, when useful signal spaces span all available dimensions.
In our approach, we relax the rank cost function to its convex envelope, the nuclear norm of interference singular values. 
Then, we approximate the full rank constraints with positivity constraints on the minimum eigenvalue of the signal space matrices restricted to be positive semidefinite, and showed that asymptotically this is a tight relaxation with respect to the RCRM objective.

\section{Interference Alignment for Cellular Networks}
In this section we consider the case of the $K$-cell interference channel as presented in \cite{Suh}, where each cell supports $d$ users. 
We tailor our algorithm for this problem by adding extra affine constraints on the entries of the precoding matrices.
Such constraints correspond to the fact that each user $u$ in a cell $k$ wishes to transmit only one symbol using a beamforming vector ${\bf v}_{k,u}\in\mathbb{C}^{\frac{M_{\text{t}}}{d}\times 1}$, where we assume $\frac{M_{\text{t}}}{d}\in\mathbb{N}^*$. 
For this system, the received signal at receiver $k$ is given by
{\small
\begin{align}
{\bf y}_k&=\sum_{u=1}^d{\bf H}^{(u)}_{k,k}{\bf v}_{k,u}x_{k,u}+\sum_{l=1,l\ne k}^K\sum_{u=1}^d{\bf H}^{(u)}_{k,l}{\bf v}_{l,u}x_{l,u}+{\bf w}_k\nonumber\\
&=\left[{\bf H}_{k,k}^{(1)}\ldots{\bf H}_{k,k}^{(d)}\right]\left[
\begin{array}{@{}c@{}c@{}c@{}}
{\bf v}_{k,1} & \ldots & {\bf 0}_{\frac{M_{\text{t}}}{d}\times1}\\
\vdots & \ddots & \vdots\\
{\bf 0}_{\frac{M_{\text{t}}}{d}\times1}& \ldots & {\bf v}_{k,d}
\end{array}
\right]\left[
\begin{array}{@{}c@{}}
x_{k,1}\\
\vdots\\
x_{k,d}
\end{array}
 \right]\\
&
+\sum_{l=1,l\ne k}^K\left[{\bf H}_{k,l}^{(1)}\ldots{\bf H}_{k,l}^{(d)}\right]\left[
\begin{array}{@{}c@{}c@{}c@{}}
{\bf v}_{l,1} & \ldots & {\bf 0}_{\frac{M_{\text{t}}}{d}\times1}\\
\vdots & \ddots & \vdots\\
{\bf 0}_{\frac{M_{\text{t}}}{d}\times1}& \ldots & {\bf v}_{l,d}
\end{array}
\right]\left[
\begin{array}{@{}c@{}}
x_{l,1}\\
\vdots\\
x_{l,d}
\end{array}
 \right]+{\bf w}_k\nonumber\\
&={\bf H}_{k,k}{\bf V}_k{\bf x}_k+\sum_{l=1,l\ne k}^K{\bf H}_{k,l}{\bf V}_l{\bf x}_l+{\bf w}_k,
\end{align}
}where ${\bf H}_{k,l}^{(u)}\in\mathbb{C}^{M_{\text{r}}\times\frac{M_{\text{t}}}{d}}$ represents the channel between user $u$ of cell $l$ and receiver $k$, ${\bf H}_{k,l}=\left[{\bf H}_{k,l}^{(1)}\ldots{\bf H}_{k,l}^{(d)}\right]\in\mathbb{C}^{M_{\text{r}}\times M_{\text{t}}}{d}$, and 
\begin{equation}
{\bf V}_k=\left[
\begin{array}{ccc}
{\bf v}_{k,1} & \ldots & {\bf 0}_{\frac{M_{\text{t}}}{d}\times1}\\
\vdots & \ddots & \vdots\\
{\bf 0}_{\frac{M_{\text{t}}}{d}\times1}& \ldots & {\bf v}_{k,d}
\end{array}
\right],
\end{equation}
where  ${\bf v}_{k,u}$ represents the beamforming vector used by user $u$ of cell $k$,for $l,k\in\mathcal{K}$.
Hence, this mathematical formulation is equivalent to a general $K$-user MIMO interference channel where symbol $x_{k,u}$ of the symbol vector ${\bf x}_k=\left[x_{k,1}\ldots x_{k,d}\right]^T$ is transmitted only from some subset of $\frac{M_{\text{t}}}{d}$ transmit antennas.

Ideally, we would like to solve $\mathcal{P}$ with the extra affine constraints corresponding to zeros at the appropriate entries of ${\bf V}_k$.
We propose the following approximation: run $\mathcal{A}(n)$, where $\mathcal{A}_{{\bf V}}\left(\left\{{\bf U}_l\right\}_{l=1}^K\right)$ has added affine constraints 
\begin{equation}
\left[{\bf V}_{k}\right]_{\mathcal{I}_u,u}={\bf 0}_{K(d-1)\times 1} \; \forall u\in\{1,2,\ldots d\},
\end{equation}
and 
\begin{equation}
\mathcal{I}_u = \left\{1,\ldots, (u-1)\frac{M_{\text{t}}}{d}\right\}\cup\left\{u\frac{M_{\text{t}}}{d}+1,\ldots,M\right\}
\end{equation}
represents the sets of indices (rows) that the precoding matrices have zero entries, for all $u\in\{1,2,\ldots,d\}$. Observe that no further approximation is made to $\mathcal{P}$ due to affine constraints being tractable.
In the simulations section we observe that the cellular case approximation scheme yields solutions with low rank interference matrices, resulting in high sum-rate.

\section{Simulations}
\subsection{Interference Channel}
In this experimental evaluation we run simulations for a $(4\times8,d=1,3)^{3}$, a $(6\times6,d=1,3)^{3}$, a $2$ time slot symbol extended single antenna, $3$-user interference channel, with $d=1$, and a $(4\times18,d=1,2)^{10}$ system.
All MIMO systems considered are proper, i.e. $d\le \frac{M_{\text{t}}+M_{\text{r}}}{K+1}$.
In our simulations, we allocate power $\frac{10^{\frac{P}{10}}}{d}$ to each column of the precoding matrices, where $P=[0:10:80]$dB, and set the noise power level to $\sigma^2=1$.
We present results averaged over $200$ channel realizations, where each channel element is drawn i.i.d. from a real Gaussian distribution with mean zero and variance $1$.
We plot the sum-rate of each system and the average number of interference free dimensions per user.
The sum rate we plot is computed as
$R = \sum_{i=1}^K\frac{1}{2}\log\det\left({\bf I}_d+\left({\bf I}_d+{\bf J}_k{\bf J}^H_k\right)^{-1}{\bf S}_k{\bf S}^H_k\right)$.
The number of interference free dimensions (for normalized input ${\bf V}_k$ and output ${\bf U}_k$ matrices) is calculated as the number of singular values of ${\bf S}_k$  with value greater than $10^{-6}$, minus the number of singular values of ${\bf J}_k$ that are greater than $10^{-6}$.
We should not that this metric does not account for the rate slope at the low SNR regime.

\subsubsection{$3$-user interference channel}
First, we consider a $(4\times8,d=1,3)^{3}$ and a $(6\times6,d=1,3)^{3}$ system.
For each simulation, we run $5$ iterations of our algorithm, $10^4$ iterations of the minimum interference leakage algorithm, and $10^4$ iterations of the max-SINR algorithm. We also calculate an orthogonalized version of the max-SINR outputs, which is denoted as max-SINR with QR in the figures.
To run $\mathcal{A}(5)$ we set $\epsilon=0.1$ and use the \texttt{CVX} toolbox \cite{cvx}. In terms of complexity, at each iteration our proposed algorithm solves $2$ semidefinite programs, the leakage minimization $2\cdot K$ eigenvalue decompositions, and the max-SINR performs $2\cdot d\cdot K$ matrix inversions. For our proposed scheme and the max-SINR with QR algorithm $2K$ QR factorizations are performed to orthogonalize the output precoding and zero-forcing matrices.

In Fig. 1 and Fig. 2, we consider a $(4\times8,d=1,3)^3$ system and plot the sum rate and average per user DoF\footnote{By a slight abuse of terminology, here we use DoF to denote the number of interference free signaling dimensions at any SNR point.} versus $P$  of the interference leakage minimization algorithm, the max-SINR approach of  \cite{Gomadam} and \cite{Peters1}, and our scheme, respectively. 
Again, for all cases considered, perfect IA is expected to be feasible with high probability since the systems are proper \cite{Yetis}.
In Fig. 1 for the $d=1$ case, we observe that at low to moderate SNRs, the max-SINR solution outperforms both our proposed algorithm and the leakage minimization with respect to the achievable data rate.
Interestingly, when we shift to higher SNRs above $40$dB, our proposed algorithm matches the performance of max-SINR (with or without QR) and both schemes offer a slight rate benefit compared to the leakage minimization approach.
For $d=3$, we observe the same trend at the low-SNR regime, where in this case both max-SINR and leakage minimization schemes achieve higher rates compared to our approach. 
However, the benefits of our algorithm become apparent for SNRs above $40$dB where the extra DoF achieved yields higher data rates compared to both the max-SINR and leakage minimization approaches.
In Fig. 2, observe that for $d=1$ both our scheme and the leakage minimization achieve exactly $d=1$ average per user DoF for all SNRs, while the max-SINR algorithm exhibits a more adaptive behavior which yields $1$ DoF at the high SNR regime, where DoF becomes an important factor of the SINR metric.
For $d=3$, both the max-SINR and leakage minimization algorithms do not achieve more than $1$ per user DoF, for $10^4$ iterations.
In this case, our nuclear norm approach seems to favor sparse solutions that yield more than $2$ DoF resulting to higher data rates.

In Fig. 3 and 4 we consider a $(6\times6,d=1,3)^3$ system.
In Fig. 3  for $d=1$, the performance of max-SINR and our proposed algorithm approximately match for SNRs greater than $40$dB and both schemes offer a constant gap rate advantage compared to the leakage minimization approach. 
For, $d=3$ the max-SINR algorithm outperforms, both our proposed algorithm and the leakage minimization approach for SNRs up to approximately $25$dB. 
For the SNR regime above $30$dB, our algorithm does not provide extra DoF for $5$ iterations, while the interference leakage provides more and yields substantially higher data rates compared to both max-SINR and our approach.
In Fig. 4 we observe that leakage minimization achieves more interference free dimensions compared to our algorithm and the max-SINR approach which results to higher achievable per user DoF.
It seems that our algorithm, for values of $d= \frac{M_{\text{t}}+M_{\text{r}}}{K+1}$, performs better when $M_{\text{t}}>M_{\text{r}}$.
Interestingly, when $d<\frac{M_{\text{t}}+M_{\text{r}}}{K+1}$ all algorithm seem to get the maximum DoF (at least in the high-SNR regime) in our simulations.

\subsubsection{$3$-user, symbol extended, interference channel}
In Fig. 5 we consider a single antenna interference channel, which we extend across $2$ time slots and plot the sum rate versus $P$  of the interference leakage minimization algorithm, the max-SINR approach, and our scheme, respectively, and set $d=1$. 
We run the interference leakage and max-SINR approaches for $10^4$ iterations and our algorithm for $5$.
For SNRs up to $25$dB the max-SINR solution gives slightly higher sum-rate compared to our proposed algorithm and the leakage minimization.
However, due to the diagonal structure  the  interference leakage scheme might generate beamforming and zero-forcing matrices that result in zero eigenvalues of signal spaces, as shown in \cite{Gomadam2}. 
These signal space deficiencies are avoided in our approach due to the positivity constraint on the minimum eigenvalue of each singnal space. 
We observe that in the high-SNR regime, our approach offers a substantial rate increase compared to max-SINR and leakage minimization approaches.

\subsubsection{$10$-user interference channel}
In Fig. 6 we consider a $(4\times18,d=1,2)^{10}$ system and plot the sum rate versus $P$  of the interference leakage minimization algorithm, the max-SINR approach and our scheme, respectively. 
Due to the size of the problem parameters, in this part, we run the interference leakage and max-SINR approaches for $2\cdot 10^3$.
For the $d=1$ case, we observe a similar trend to the $3$-user system: at the low to moderate SNR regime the max-SINR solution outperforms both our proposed algorithm and the leakage minimization with respect to the achievable data rate. Then, the rate performance of the max-SINR and our approach seem to match and be slightly higher compared to the rate achieved by the leakage minimization.
For $d=2$, we observe that up to the SNR of $40$dB the max-SINR achieves higher data rate compared to our approach and the leakage minimization.
Past the $40$dB mark, the leakage minimization and max-SINR with QR exhibit a zero slope sum-rate curve.
Eventually, it seems that the extra DoF obtained by our algorithm here give slightly better performance compared to the max-SINR at the high-SNR regime.

\subsection{Cellular Interference Channel}

In Fig. 7, we consider a $3$-cell interference channel, where each cell has $2$ users. Each user is equipped with $3$ antennas, and each receiver has  $4$ receive antennas. 
For this system, we compare our algorithm with a random beamforming (BF) and interference zeroforcing scheme.
In the latter scheme, the users randomly select beamforming vectors and the receivers use zeroforcing matrices with columns the $d$ eigenvectors associated with the $d$ smallest eigenvalues of the interference correlation matrix, like the one defined in (\ref{QI}).
We run these two scheme for $200$ channel realizations and plot the data rate versus $P=[0:10:60]$dB. 
Our proposed algorithm runs for $1$, $2$ and $10$ iterations. 
For $1$ iteration our scheme's performance is comparable with the random BF and interference zeroforcing.
Interestingly, we observe that there is a substantial increase in data rate for $2$ iterations (approximately $2$ times more sum-rate at $60$dB) and $10$ iterations (approximately $3$ times more sum-rate at $60$dB), where our algorithm outperforms random BF and interference zero-forcing for all power configurations.

\section{Conclusion}

To conclude, in this work we reformulated the interference alignment problem as a rank constrained, rank minimization.
This framework allowed us to introduce individually tight convex relaxations for the cost function and constraints of the RCRM problem. 
Our heuristic was inspired by the nuclear norm relaxation of rank introduced in \cite{Recht}, however, in this paper we did not establish theoretical guarantees under which this relaxation is tight.
Such a theoretical investigation would be a very interesting open problem for future research.

\newpage

\begin{figure}[t]
\includegraphics[width=1\columnwidth]{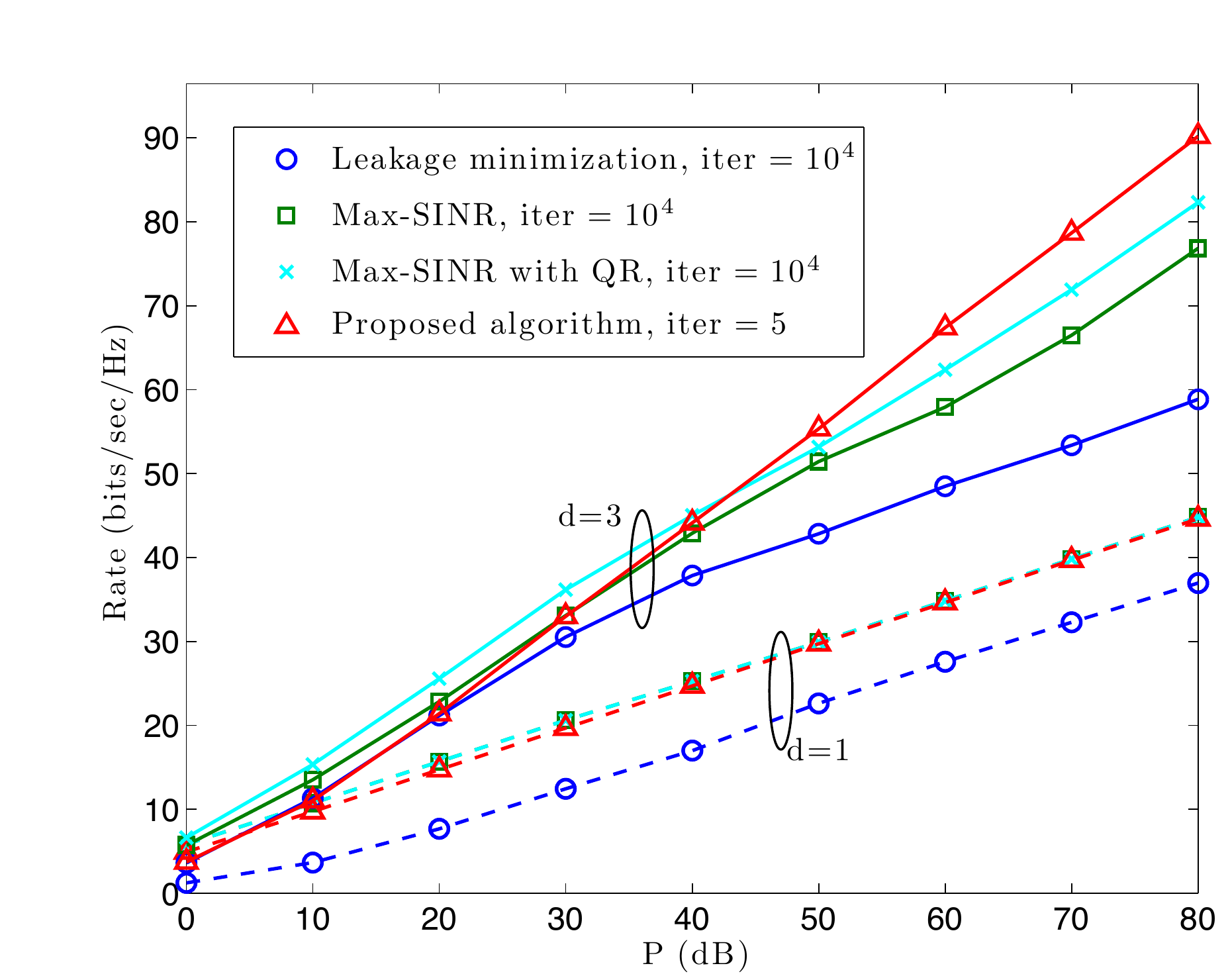}
\caption{Sum-rate vs. $P$, for a $3$-user system, $M_{\text{r}}=4$, $M_{\text{t}}=8$, and $d=1,3$.}
\end{figure}

\newpage

\begin{figure}[t]
\includegraphics[width=1\columnwidth]{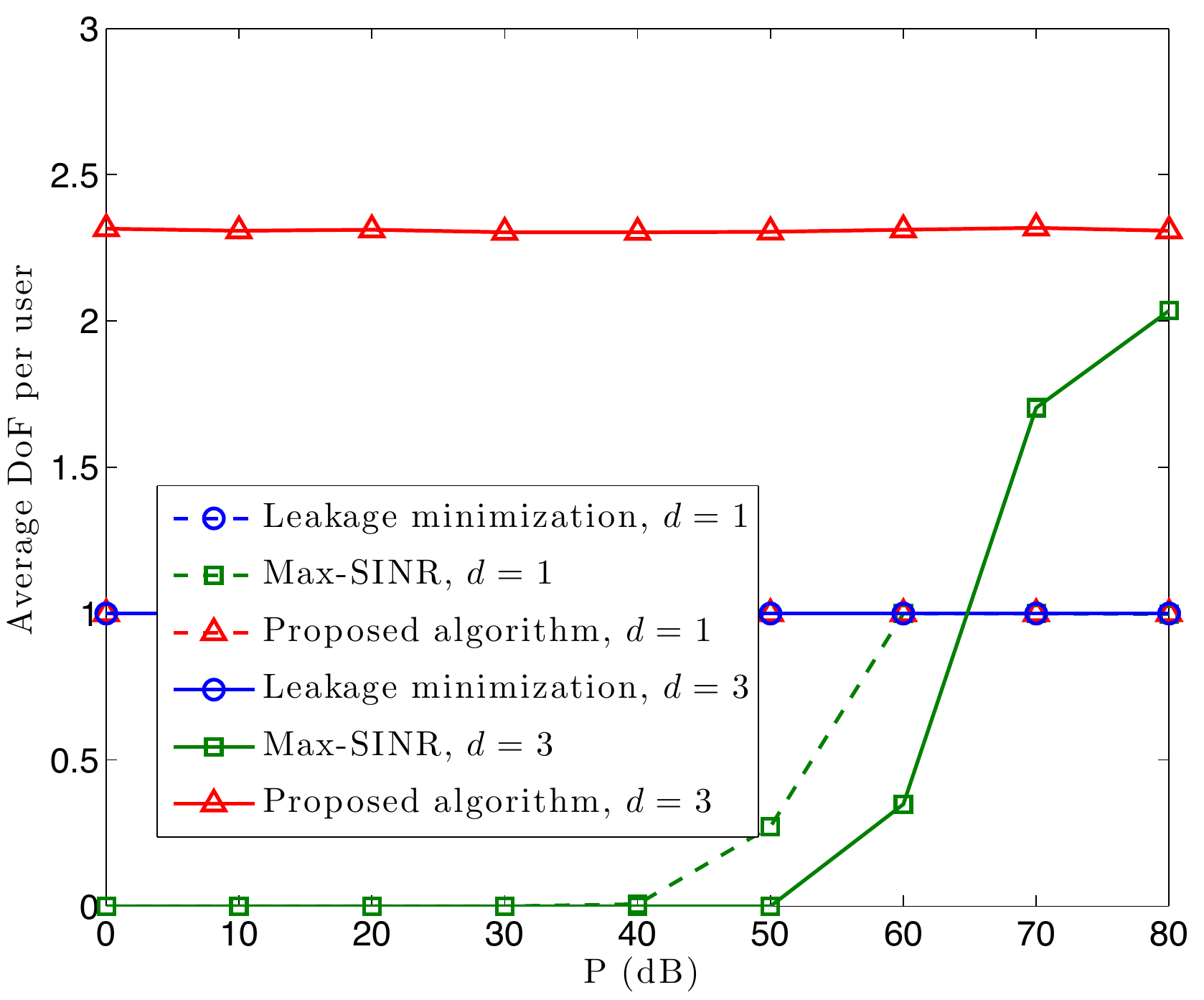}
\caption{Average number of interference free dimensions per user vs. $P$, for a $3$-user system, $M_{\text{r}}=4$, $M_{\text{t}}=8$, and $d=1,3$.}
\end{figure}

\newpage 

\begin{figure}[t]
\includegraphics[width=1\columnwidth]{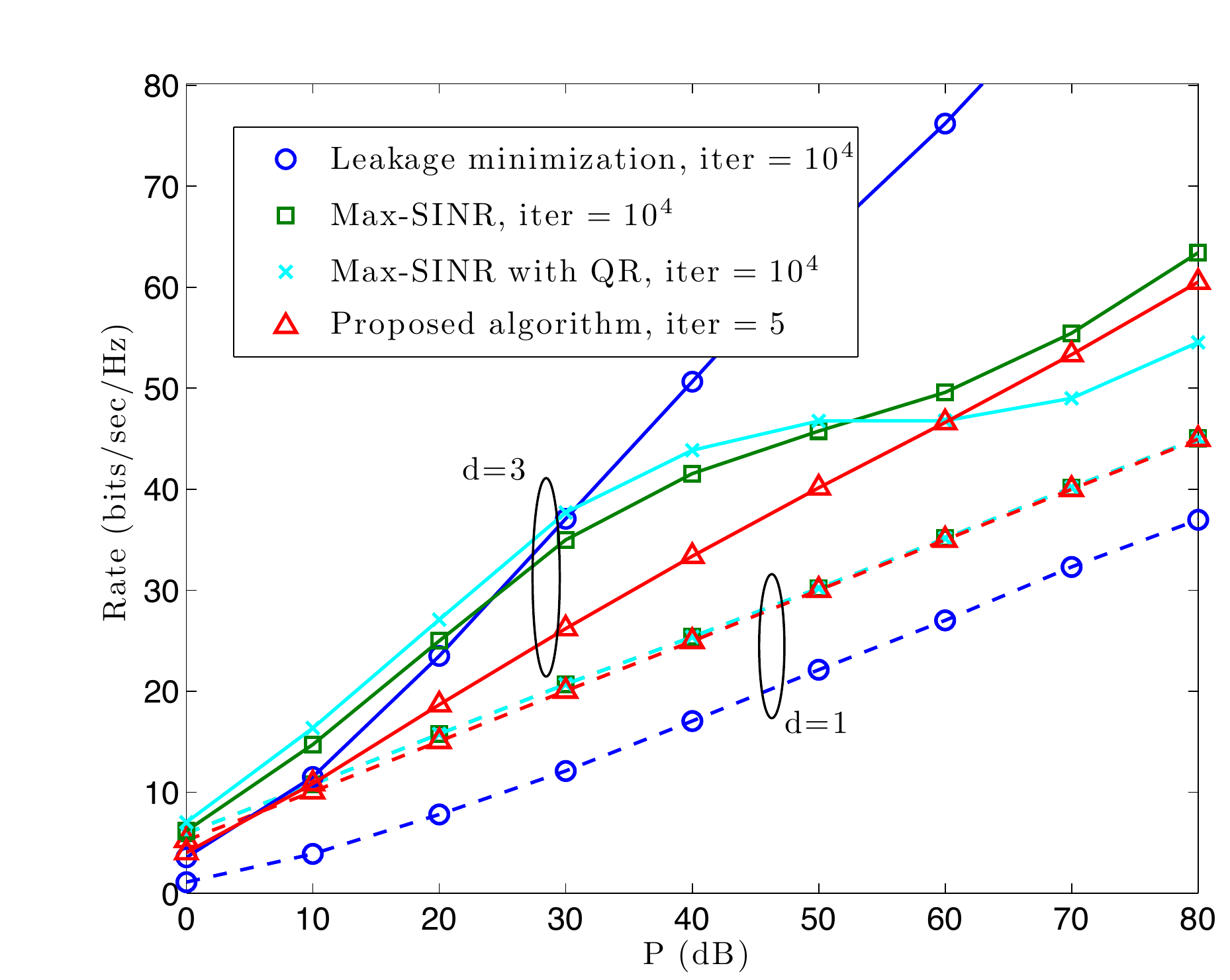}
\caption{Sum-rate vs. $P$, for a $3$-user system, $M_{\text{r}}=6$, $M_{\text{t}}=6$, and $d=1,3$.}
\end{figure}

\newpage 
\begin{figure}[t]
\includegraphics[width=1\columnwidth]{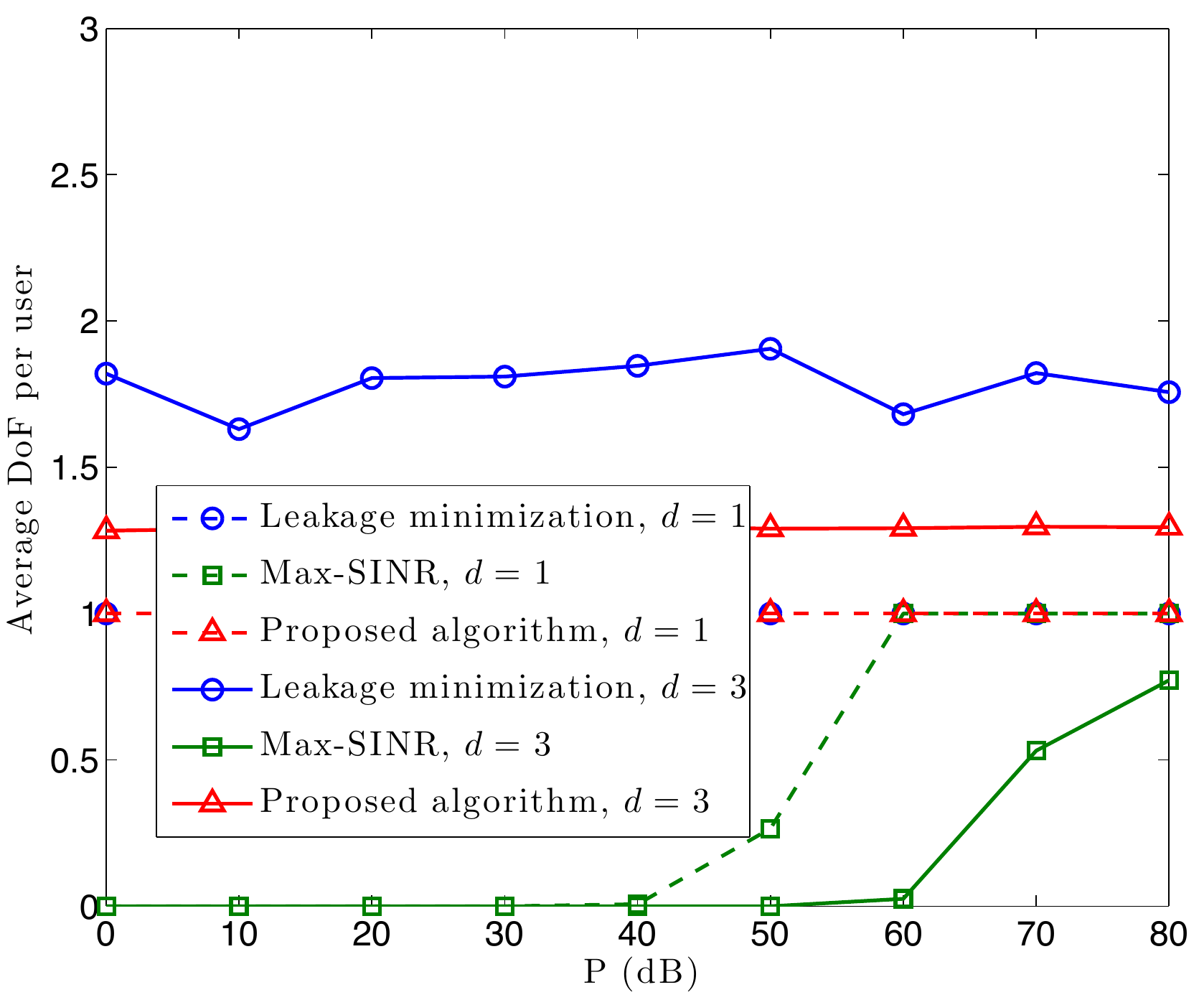}
\caption{Average number of interference free dimensions per user vs. $P$, for a $3$-user system, $M_{\text{r}}=6$, $M_{\text{t}}=6$, and $d=1,3$.}
\end{figure}

\newpage 

\begin{figure}[t]
\includegraphics[width=1\columnwidth]{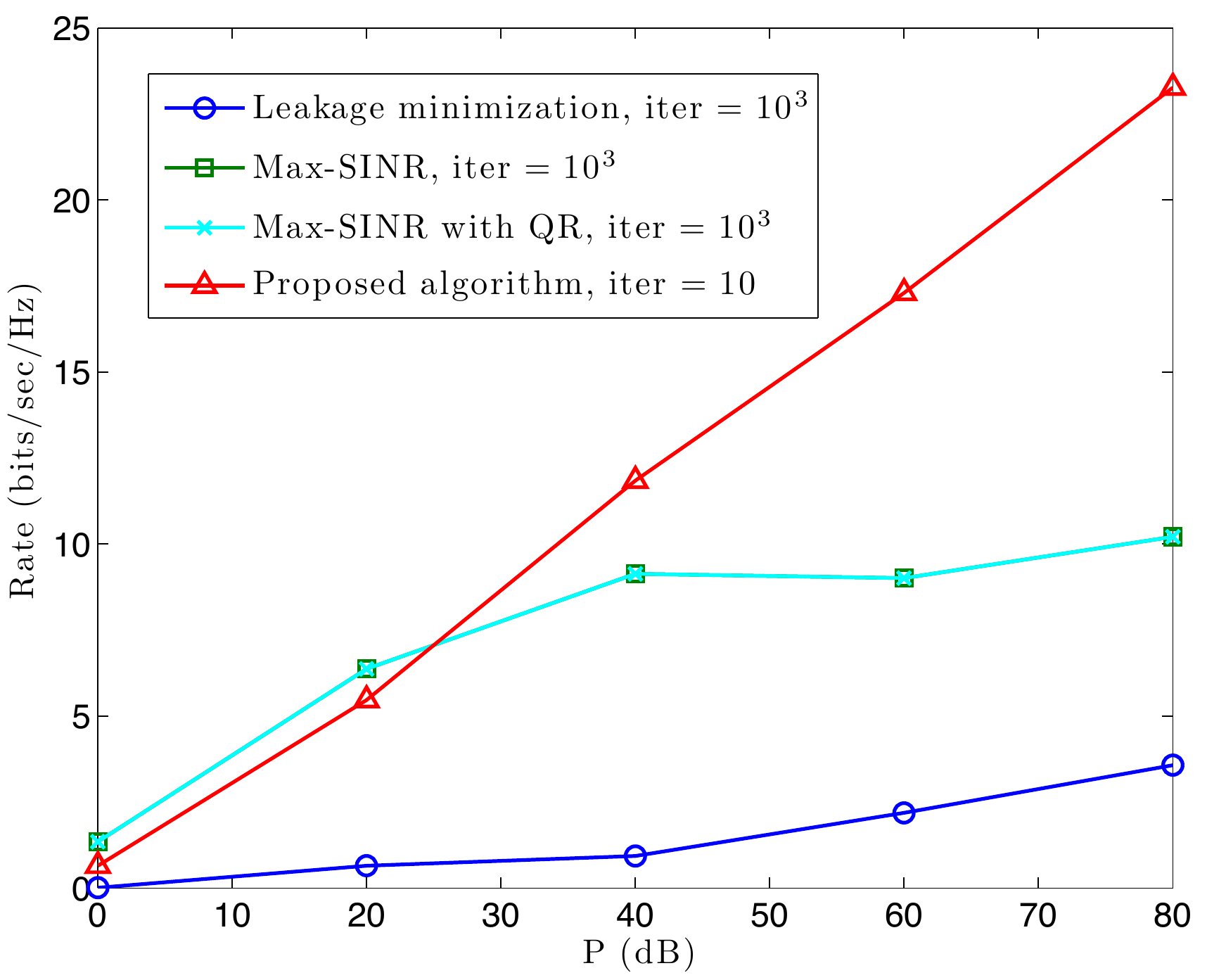}
\caption{Sum-rate vs. $P$, for a $3$-user system, single antenna system using a length $2$ symbol extension, where $d=1$.}
\end{figure}

\newpage

\begin{figure}[t]
\centerline{\includegraphics[width=1\columnwidth]{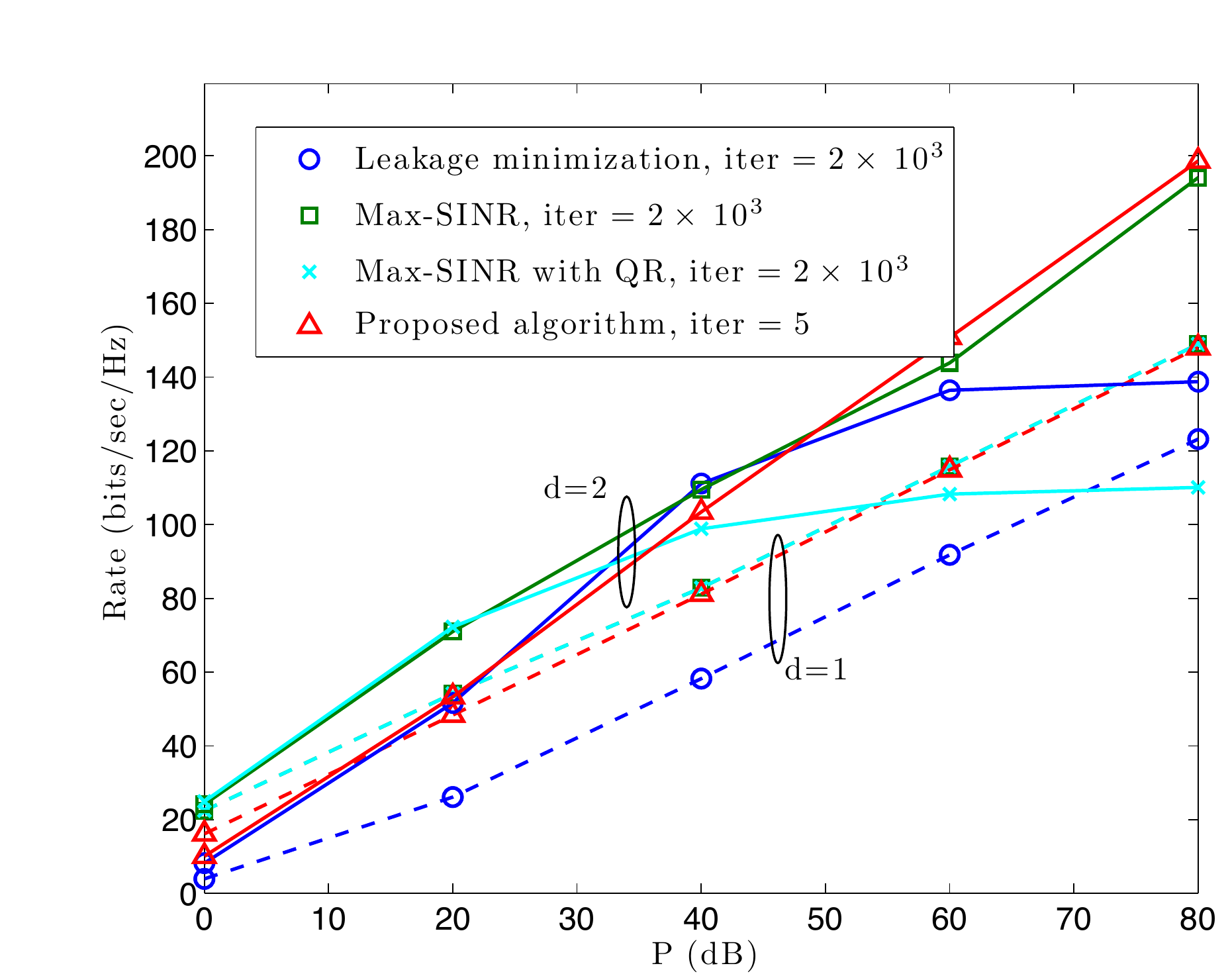}}
\caption{Sum-rate vs. $P$, for a $10$-user system, $M_{\text{r}}=4$, $M_{\text{t}}=18$, and $d=1,2$.}
\end{figure}

\begin{figure}[t]
\includegraphics[width=1\columnwidth]{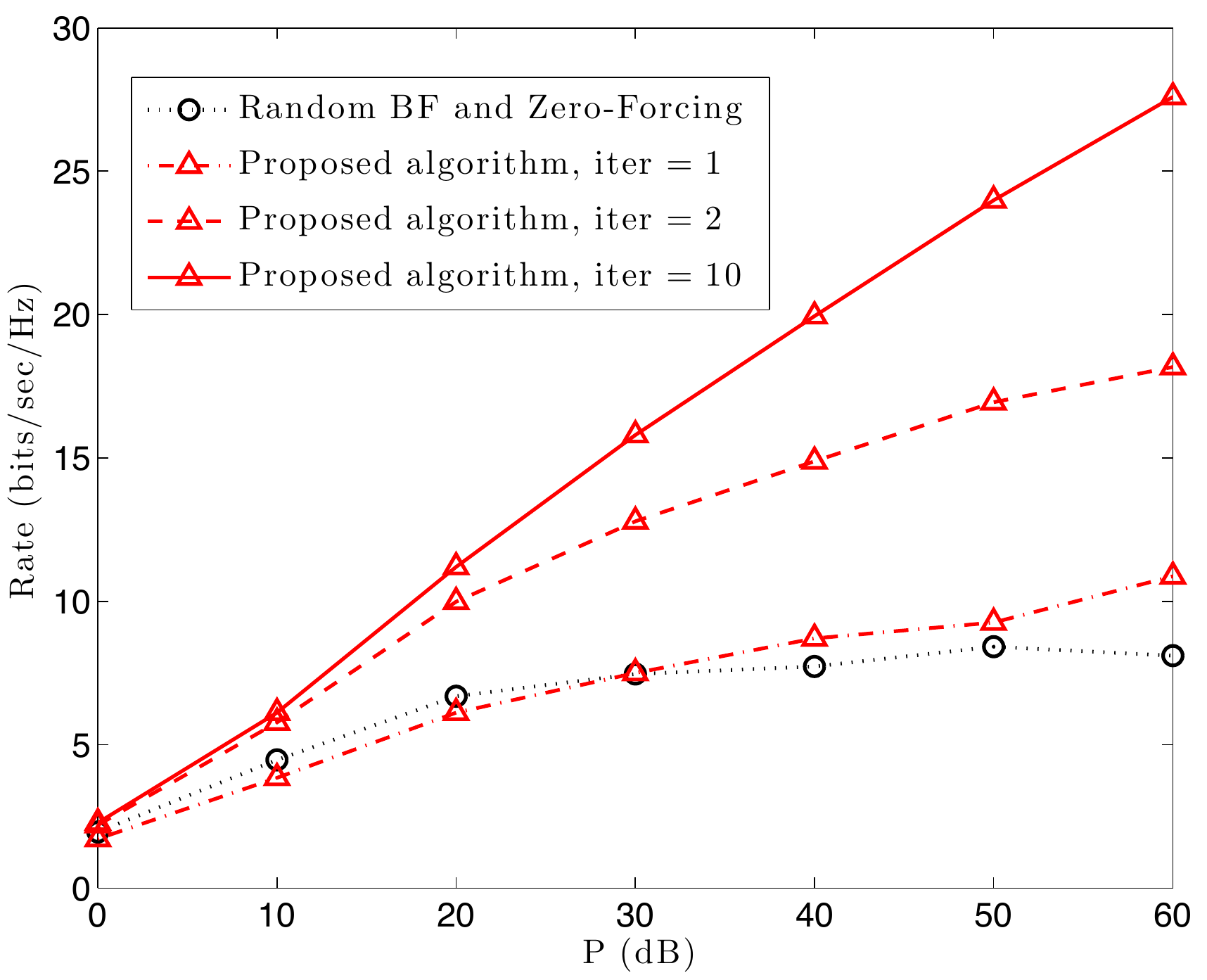}
\caption{Sum-rate vs. $P$, for a $3$-cell interference channel, with $2$ users per cell, $3$ transmit antennas per user, and $4$ antennas per receiver.}
\end{figure}


\begin{thebibliography}{99}
\renewcommand{\baselinestretch}{1}
\footnotesize

\bibitem{PD}
D. S. Papailiopoulos and A. G. Dimakis, ``Interference alignment as a rank
  constrained rank minimization,'' in {\em IEEE GLOBECOM 2010}, Miami, FL,
  USA.

\bibitem{Cadambe1}
V. R. Cadambe  and S. A. Jafar,
``Interference alignment and spatial degrees of freedom for the $K$ user interference channel,''
{\it IEEE Trans. on Inform. Theory},
vol. 54, 
pp. 3425-3441, 
Aug. 2008.,

\bibitem{Cadambe3}
V. R. Cadambe and S. A. Jafar, 
``Reflections on interference alignment and the degrees of freedom of the $K$ user interference channel,''
{\it IEEE Inform. Theory Soc. Newsletter}, 
vol. 59, 
pp. 5-9,
Dec. 2009.

\bibitem{Ali1}
M. M.-Ali, A. Motahari, and A. Khandani, 
``Signaling over MIMO multi-base systems - combination of multi-access and broadcast schemes,''
in {\it Proc. of ISIT 2006},
pp. 2104-2108,
2006.

\bibitem{Jafar1}
S. Jafar and M. Fakhereddin,
``Degrees of freedom for the mimo interference channel,'' 
{\it IEEE Trans. on Inform. Theory}, 
vol. 53, 
pp. 2637-2642, 
July 2007.

\bibitem{Jafar2}
S. Jafar and S. Shamai, 
``Degrees of freedom region for the MIMO $X$ channel,'' 
{\it IEEE Trans. on Inform. Theory,} 
vol. 54, 
pp. 151-170, 
Jan. 2008.


\bibitem{Ali2}
M. M.-Ali, A. Motahari, and A. Khandani, 
``Communication over MIMO $X$ channels: Interference alignment, decomposition, and performance analysis,'' 
{\it in IEEE Trans. on Inform. Theory}, 
pp. 3457-3470.
Aug. 2008.


\bibitem{Gou}
T. Gou and S. Jafar, 
``Degrees of freedom of the $K$ user MIMO interference channel,'' 
in {\it 42nd Asilomar Conf. on Signals, Systems and Computers 2008},
Oct. 2008.

\bibitem{Gomadam}
K. Gomadam, V. Cadambe, and S. Jafar, 
``Approaching the capacity of wireless networks through distributed interference alignment,''
 in {\it Proceedings of IEEE GLOBECOM 2008}, 
Dec. 2008.


\bibitem{Yetis} 
C. Yetis, T. Gou, S. Jafar, and A. Kayran, 
``Feasibility conditions for interference alignment,''
 in {\it Proceedings of IEEE GLOBECOM 2009}, 
Nov. 2009.


\bibitem{Suh}
C. Suh and D. Tse,
``Interference alignment for cellular networks,''
in {\it Proceedings of 40th Annual Allerton Conf. on Commun., Control, and Computing 2008},
Sep. 2008.

\bibitem{Roland}
R. Tresch, M. Guillaud, and E. Riegler,
``On the achievability of interference alignment in the $K$-user constant MIMO interference channel,''
arXiv.org:0904.4343v1 [cs.IT],
2009.

\bibitem{Luo}
M. Razaviyayn, M. Sanjabi, and Z.-Q. Luo
``Linear transceiver design for interference alignment: Complexity and computation,''
arXiv:1009.3481v1 [cs.IT],
Sep. 2010.

\bibitem{Peters1}
 S. W. Peters and R. W. Heath, Jr., 
``Interference alignment via alternating minimization,''
 in {\it Proc. IEEE ICASSP 2009}, 
 pp. 2445-2448,
Apr. 2009.

\bibitem{Peters2}
S. W. Peters and R. W. Heath Jr.,
``Cooperative algorithms for MIMO interference channels,''
arXiv:1002.0424v2 [cs.IT],
Oct. 2010.

\bibitem{Gomadam2}
K. Gomadam, V. Cadambe, and S. Jafar, 
``A distributed numerical approach to interference alignment and applications to wireless interference networks,''
 to appear in {\it IEEE Trans. Inform. Theory}.

\bibitem{Sung1}
H. Yu, J. Park, Y. Sung, and Y. H. Lee,
``A least squares approach to joint beam design for interference alignment in multiuser interference channels,''
{\it IEEE Trans. on Signal Proc.},
vol. 58, 
pp. 4960 - 4966,
Sept. 2010.

\bibitem{Sung2}
J. Park, Y. Sung, H. V. Poor,
``On beamformer design for multiuser MIMO interference channels,''
arXiv:1011.6121v1 [cs.IT],
Nov. 2010.


\bibitem{Recht}
B. Recht, M. Fazel, and P. Parrilo,
``Guaranteed minimum rank solutions of matrix equations via nuclear norm minimization,'' 
{\it Submitted to SIAM Review}, 
2007.

\bibitem{Candes}
E. J. Cand\`{e}s and T. Tao,
``The power of convex relaxation: Near-optimal matrix completion,''
{\it IEEE Trans. Inform. Theory} (to appear).

\bibitem{cvx}
M. Grant and S. Boyd,
CVX: Matlab software for disciplined convex programming (web page and software),
http://stanford.edu/$\sim$boyd/cvx, 
June 2009.





\end{thebibliography}
\end{document}